# Superflares and Giant Planets

*From time to time, a few sunlike stars produce gargantuan outbursts. Large planets in tight orbits might account for these eruptions*

Eric P. Rubenstein

Envision a pale blue planet, not unlike the Earth, orbiting a yellow star in some distant corner of the Galaxy. This exercise need not challenge the imagination. After all, astronomers have now uncovered some 50 "extrasolar" planets (albeit giant ones). Now suppose for a moment something less likely: that this planet teems with life and is, perhaps, populated by intelligent beings, ones who enjoy looking up at the sky from time to time.

During the day, these creatures would see their sun shining brightly, providing them with an almost constant bath of warming rays. At night they might gaze at a moon or two against a backdrop of seemingly unchanging stars. Over the billions of years since life first evolved on this hypothetical planet, this world has been spared any cosmic catastrophes. No nearby supernovae have exploded. The orbit of the planet has stayed circular. And no great asteroids have come crashing down. In short, none of the many celestial events that would threaten life here have happened. But today is a bad day.

A stellar flare, enormous by normal standards, erupts from the surface of the parent star, giving off a sudden burst of radiation and spewing into space a fusillade of charged particles. Although it will startle the inhabitants, the sudden heat wave will not cause bushes to burst into flames. Nor will the surface of the planet feel the blast of ultraviolet light and x rays, which will be absorbed high in the atmosphere. But the more energetic component of these x rays and the charged particles that follow them are going to create havoc when they strike air molecules and trigger the production of nitrogen oxides, which rapidly destroy ozone.

So in the space of a few days the protective blanket of ozone around this planet will largely disintegrate, allowing ultraviolet light to rain down unimpeded. Until this shield is restored, a process that might take decades, daylight will remain deadly to the organisms exposed to it.

Fortunately, our Sun does not appear prone to such disastrous explosions. Indeed, it produces a remarkably steady output of energy—the value of which is duly named the solar constant. Sensitive instruments placed on satellites show that it varies by only a few parts in a thousand. Astrophysicists do, however, surmise that the solar constant is slowly increasing, by about one percent each 300 million years. Over the 4.6 billion years since the solar system formed, the Sun has brightened by about 40 percent. Although this variation is substantial, it has taken place so gradually that changes in the amount of heat-trapping gases in the atmosphere have compensated, allowing the temperature of the Earth to remain relatively steady—and in a range that keeps water liquid—for billions of years.

Life on Earth clearly benefits from this long-term stability and from the Sun having only moderate levels of what astronomers refer to as "activity," meaning violent outbursts. In large part, this activity manifests itself as solar flares, which typically last a fraction of an hour and release their energy in a combination of charged particles, ultraviolet light and x rays. Thankfully, this radiation does not reach dangerous levels at the surface of the Earth: The terrestrial magnetic field easily deflects the charged particles, the upper atmosphere screens out the x rays, and the stratospheric ozone layer absorbs most of the ultraviolet light. So solar flares, even the largest ones, normally pass uneventfully. People living at high latitudes may see unusually intense auroral displays. But unless an especially strong flare knocks out a communication satellite or a power grid, few people take much notice.

**Inconstant Suns**

Not all stars shine as steadily as the Sun. Some change in brightness over a period of hours, days or months. Astronomers are fascinated by such *variable stars*, and groups of observers have organized specifically to examine them. Fortunately for those of us interested in such curious stars, there is an uncountable number of them.

Astronomers classify these variable stars according to their properties. So-called *geometric variables* do not actually change the amount of energy they give off. Rather, the geometry of viewing these multiple star systems causes us to see different amounts of light at different times. In most cases the variation arises because one star in a binary system passes in front of the other. Such eclipses are most likely to be seen if the two stars revolve around each other in a tight orbit. Interestingly, such binaries, ones with small separations between the stars, commonly exhibit additional kinds of variability. Tidal interactions cause these closely spaced stars to spin rapidly, which in turn greatly enhances their magnetic fields, causing strong stellar flares. Astronomers call such flaring *chromospheric activity*, because the energy comes from the portion of the stellar atmosphere that also gives the star its characteristic color.

*Eric P. Rubenstein received a Ph.D. in astronomy from Yale University in 1997. He remained there for two years before becoming a National Science Foundation International Research Fellow and starting work at the Cerro Tololo Inter-American Observatory in Chile. Last summer he returned to Yale as a lecturer in the Department of Astronomy. Address: Yale University, Department of Astronomy, P.O. Box 208101, New Haven, CT 06520-8101. Internet: ericr@astro.yale.edu*

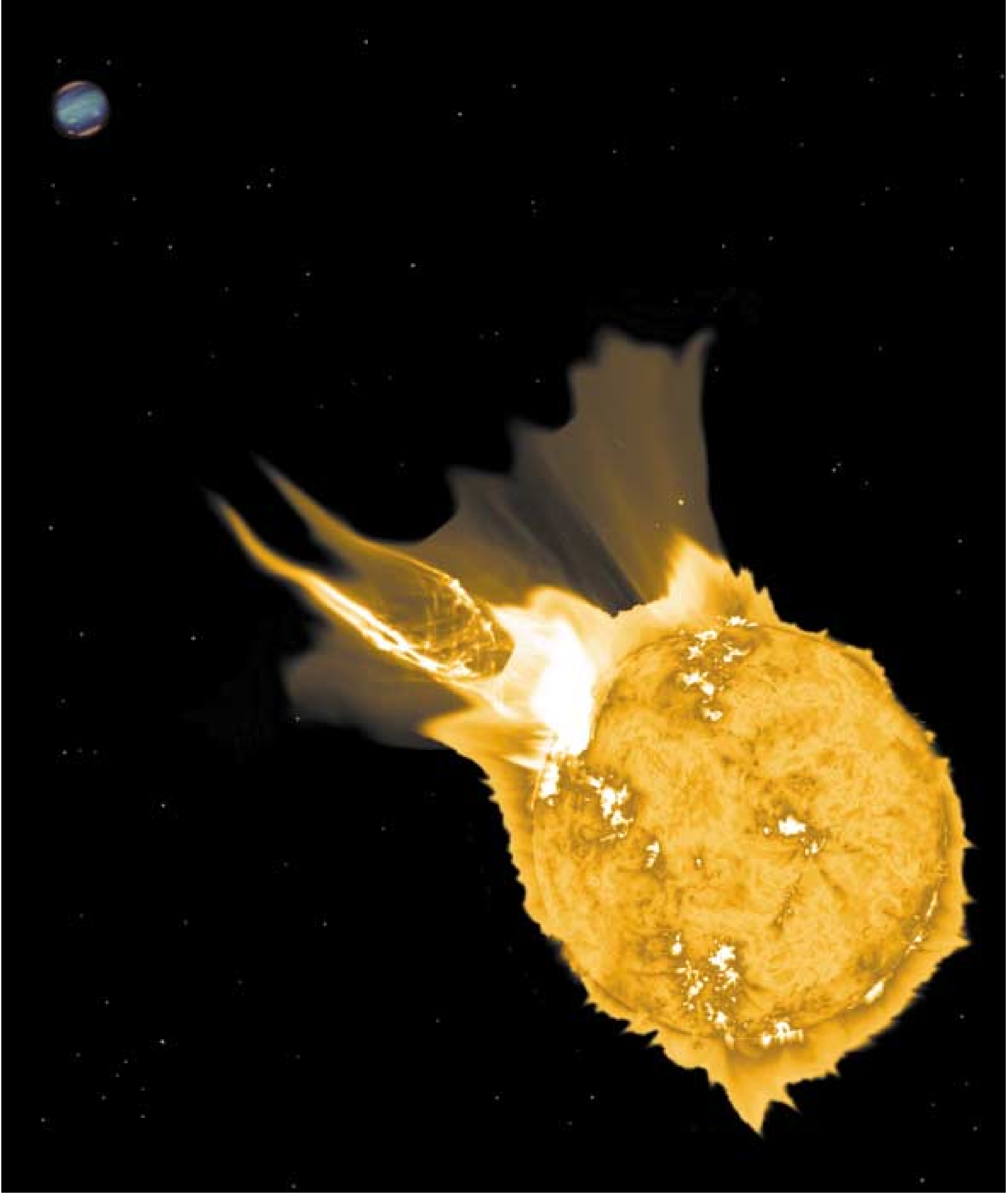

**Figure 1. Artistic rendition of a distant, sunlike star undergoing a "superflare" hints at the enormity of such an event. These huge eruptions dwarf the largest flares on the Sun, releasing from 100 to 10 million times as much energy. The author proposes that giant planets orbiting close to such stars trigger these occasional outpourings. (Image courtesy of the author.)**

Even more spectacular are the *eruptive variables*, which include supernovae—stars that end their lives by exploding so violently that they can outshine all the other stars in their galaxy. Other examples of this class include binary star systems in which a compact white dwarf pulls in hydrogen from its nearby partner. From time to time the rain of gas gives rise to huge thermonuclear explosions on the surface of the white dwarf. Such systems, dubbed *cataclysmic variables*, typically experience repeated explosions, so the term "cataclysmic" is not really apt.

Skywatchers also recognize *pulsating variables*, which alternately increase and decrease in size over intervals that range from several seconds to hundreds of days. Some vary in brightness by more than an order of magnitude. Astrophysicists have determined that such changes result from instabilities in either the rate of energy generation or in the relationship between temperature and opacity in the stellar atmosphere. Pulsators are usually periodic, although some have multiple periodicities superposed.

So variable stars are themselves a varied lot—and a well-studied one. Indeed, astronomers have prided themselves in their understanding of the mechanisms that bring about these changes. Conversely, we generally believed that we could predict which stars would be as stable as the Sun. We have traditionally assumed, for instance, that if a star has roughly the same surface temperature and luminosity as the Sun, is a single star and rotates at a speed similar to that of the Sun, it will likewise have only modest levels of chromospheric activity. Such stars are commonly called *solar analogues*. The unspoken assumption that all solar analogues are, in essence, interchangeable underlies much of the thinking about habitable worlds, and perhaps life, existing elsewhere in the cosmos.

Although it seems reasonable that solar analogues should be as stable as the Sun, it is clear that this assumption was premature—and wrong. Astronomers now know of nine instances of these sunlike stars having eruptions that were from 100 to 10 million times more energetic than the most powerful solar flares. In these cases the stars temporarily brightened by amounts that range from several percent up to a factor of 10 or more.

Startling as they would be for someone living nearby, giant flares around distant stars can easily elude us earthlings. For the most part, the sightings of these events were simply fortuitous. One set of observations was, for example, made on March 6, 1899, when a newly discovered comet passed through the constellation Fornax and entered the part of the sky that contains a star called S Fornacis. Four experienced and highly respected observers went out early that evening to view the comet. Three of them (one in Austria, one in Italy and one in Germany) noted that S Fornacis was strangely bright—estimating by eye that it was some 15 times as luminous as normal. These three sets of observations were made during one 17-minute interval. Less than a half-hour earlier, another trained astronomer in Italy had observed S Fornacis and the comet, and he did not comment on any unusual brightening. A photograph taken at Harvard Observatory a few hours later fails to show S Fornacis as anything other than normal.

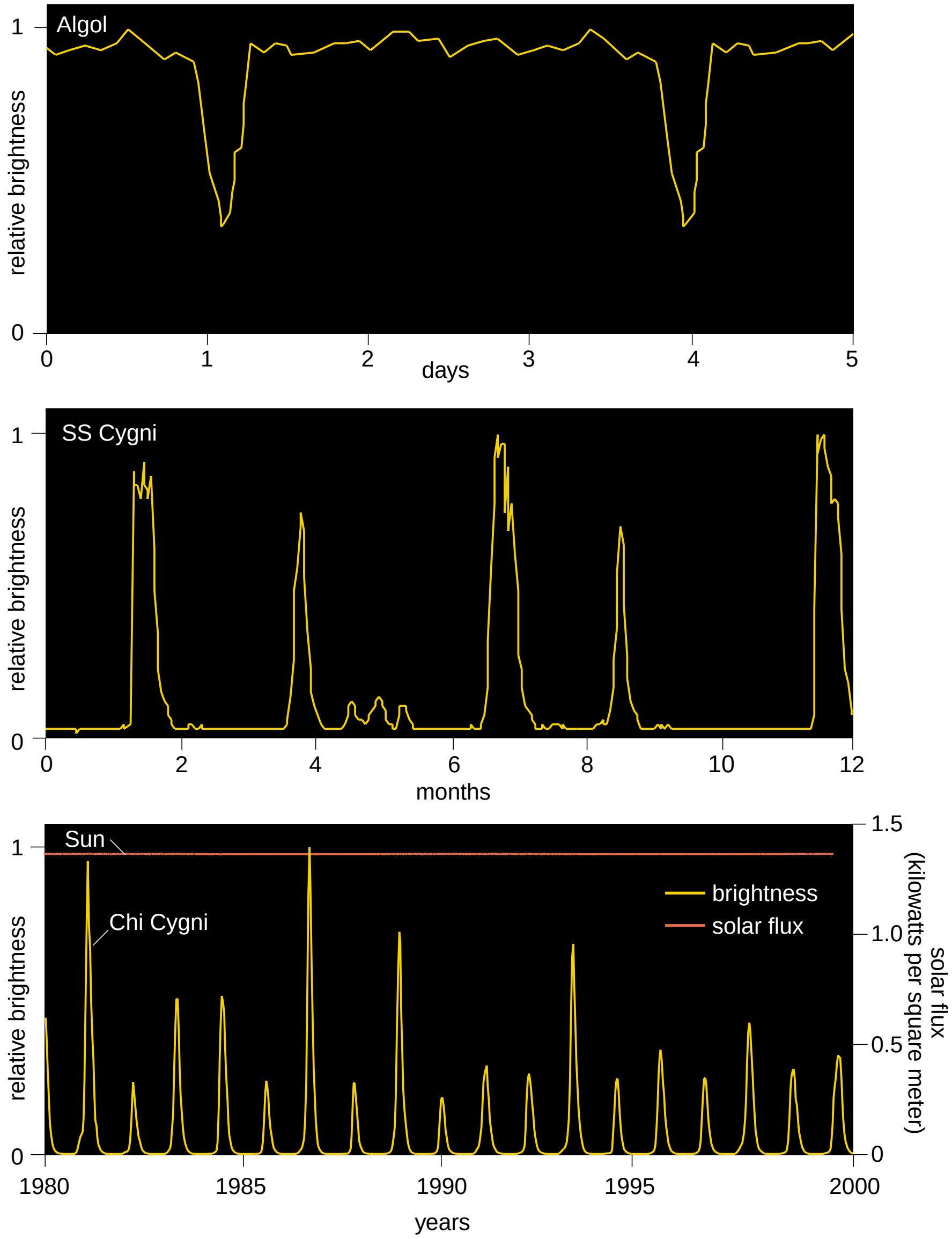

**Figure 2. Many variable stars exhibit characteristic patterns of changing brightness, which can reveal much about their nature. The star Algol (Beta Persei), for example, shows a repetitive pattern of short-lived diminutions in brightness *(top)*. Astronomers regard Algol as a "geometric variable" because the changes in brightness result from the viewing geometry, which allows each star in this binary system to periodically eclipse the other. The star SS Cygni, an "eruptive variable," shows repeated flashes, but the interval between bursts is not steady *(middle)*. Another variable star in the same constellation is Chi Cygni, which is considered a "pulsator." It changes dramatically in brightness with a regular period *(bottom)*. Solar output for the same interval shows so little variation that it appears essentially constant. (Measurements courtesy of the American Association of Variable Star Observers.)**

Although these century-old observations of an anomalous brightening are subject to the limits of human perception, the combination of three widely separated observers all seeing a dramatic change suggests that the star was, in fact, unusually luminous for a short time. The reports become even more credible when compared with some later observations of similar episodes. A series of photographs taken at the Allegheny Observatory in 1939, for example, shows a star called Groombridge 1830, another solar analogue, becoming almost twice as luminous as normal for a little longer than 18 minutes. More recently, satellites equipped with x-ray telescopes have captured two sunlike stars in the act of giving off such massive flares that they, too, brighten substantially for a few minutes to hours. And astronomers know of five other sunlike stars that became strangely luminous for a while. Although any one particular observation may be somehow defective or misleading, collectively these nine separate reports indicate that not all sunlike stars are truly sunlike.

Brad Schaefer, my former colleague at Yale (who is now at the University of Texas at Austin), has been investigating these objects for more then 10 years. His 1989 paper discussing these events introduced the idea that these stars had all experienced what he called “superflares.” But at that time he had not yet collected all the observational evidence needed to demonstrate that these sudden blow-ups did not come from some strange type of binary star or from young single stars displaying expected amounts of chromospheric activity. Recently Schaefer, with Jeremy King of the Space Telescope Science Institute and Constantine Deliyannis of Indiana University, revisited these objects and found that they are indeed normal, middle-aged stars: They are not closely spaced binaries, and they are not spinning as fast as young stars do. This later point is important because rapid rotation, even of single stars, boosts chromospheric activity

| name | rotation rate (kilometers per second) | duration | energy released (ergs) |
|---|---|---|---|
| Groombridge 1830 | 1.3 | 18 minutes | $1 \times 10^{35}$ |
| Kappa Ceti | 8 | 40 minutes | $2 \times 10^{34}$ |
| MT Tauri | ? | 10 minutes | $1 \times 10^{35}$ |
| Pi$^{1}$ Ursae Majoris | 9.7 | > 35 minutes | $2 \times 10^{33}$ |
| S Fornacis | 7 | 17–367 minutes | $2 \times 10^{38}$ |
| BD + 10°2783 | 4 | 49 minutes | $3 \times 10^{34}$ |
| Omicron Aquilae | 3.9 | 5–15 days | $9 \times 10^{37}$ |
| 5 Serpentis | 2 | 3–25 days | $7 \times 10^{37}$ |
| UU Corona Borealis | 6 | > 57 minutes | $7 \times 10^{35}$ |

**Figure 3. Nine stars are known to have experienced at least one superflare. The values given here for the amount of energy released are minimum estimates, because radiation from these events was registered for only a limited range of wavelengths. Rotation rates shown are also minimum values, because of uncertainties in the viewing geometry.**

and often sparks large-scale flaring and the emission of x rays.

Just how fast are they spinning? Astronomers often estimate how quickly objects move using Doppler shifts in the frequency of spectral lines. Just as a whistle of a passing locomotive changes in tone depending on the speed of the train toward or away from the observer, changes in spectral lines can reveal motion on the surface of a distant star. But it is difficult for astronomers to characterize stellar rotation with a single number, because stars are fluid enough that different parts rotate at different rates. So we usually just report the average speed of motion on the face of the star.

The fastest measured rotation speed of the nine known “superflaring” stars is a bit under 10 kilometers per second. Young, active stars rotate much faster. One possibility is that these indications of slow rotation are misleading. If the spin axis of the star is tilted toward the line of sight, the estimated rotation velocity will be smaller than the actual rate. So the number determined using this strategy represents a lower bound and need not reflect the true rotation rate. In an extreme case, where the spin axis of the star is in the line of sight, the measured rotation velocity will seemingly be zero, even with the star spinning rapidly.

Such unfavorable geometry might account for the low rotation rates found for one or two of these superflaring stars—but it is unlikely to explain all nine. And some indirect evidence also indicates that these stars are not spinning very quickly. The argument goes like this: Rapidly rotating stars usually contain a lot of lithium, a rather fragile element that is destroyed when it gets mixed into a hot stellar interior. Rapid rotation is thought to prevent such mixing. So by estimating the abun-dance of lithium, astronomers can gauge the rotation rate of a star without knowing how the spin axis is tilted. The nine superflaring stars all have low lithium values, which confirms that they are indeed spinning comparatively slowly.

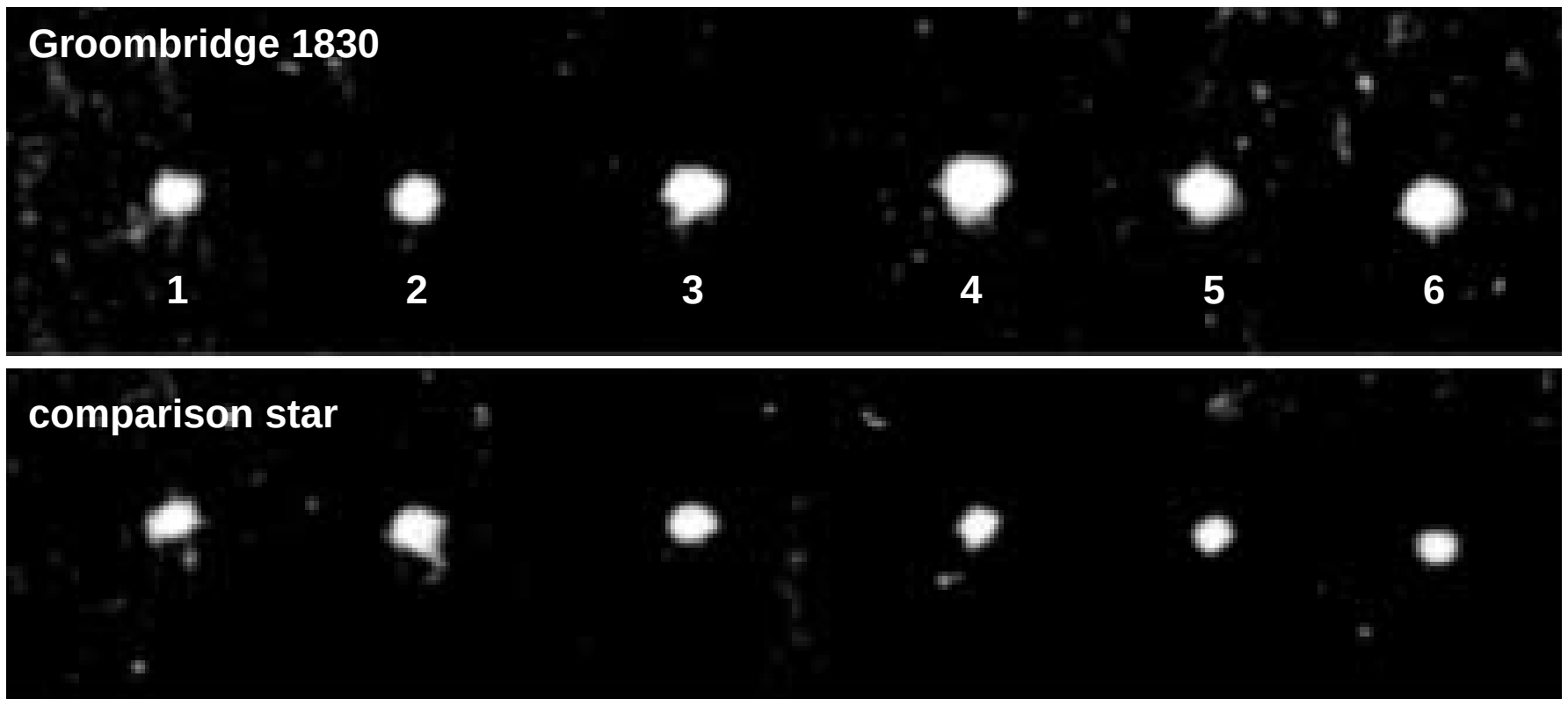

**Figure 4. Groombridge 1830, a star about 29 light-years away, experienced a superflare that made it noticeably brighter for a brief time. By happenstance, this transient event was captured by workers at the Allegheny Observatory in 1939. The six images of the star *(top)* were taken consecutively with each exposed for two minutes. Ten minutes separate image 3 from image 4, which shows an obvious increase in brightness. A comparison star captured on the same photographic plate *(bottom)* displays no such variation, confirming that the temporary brightening seen in image 4 is not an artifact. (After Beardsley *et al.*, 1974.)**

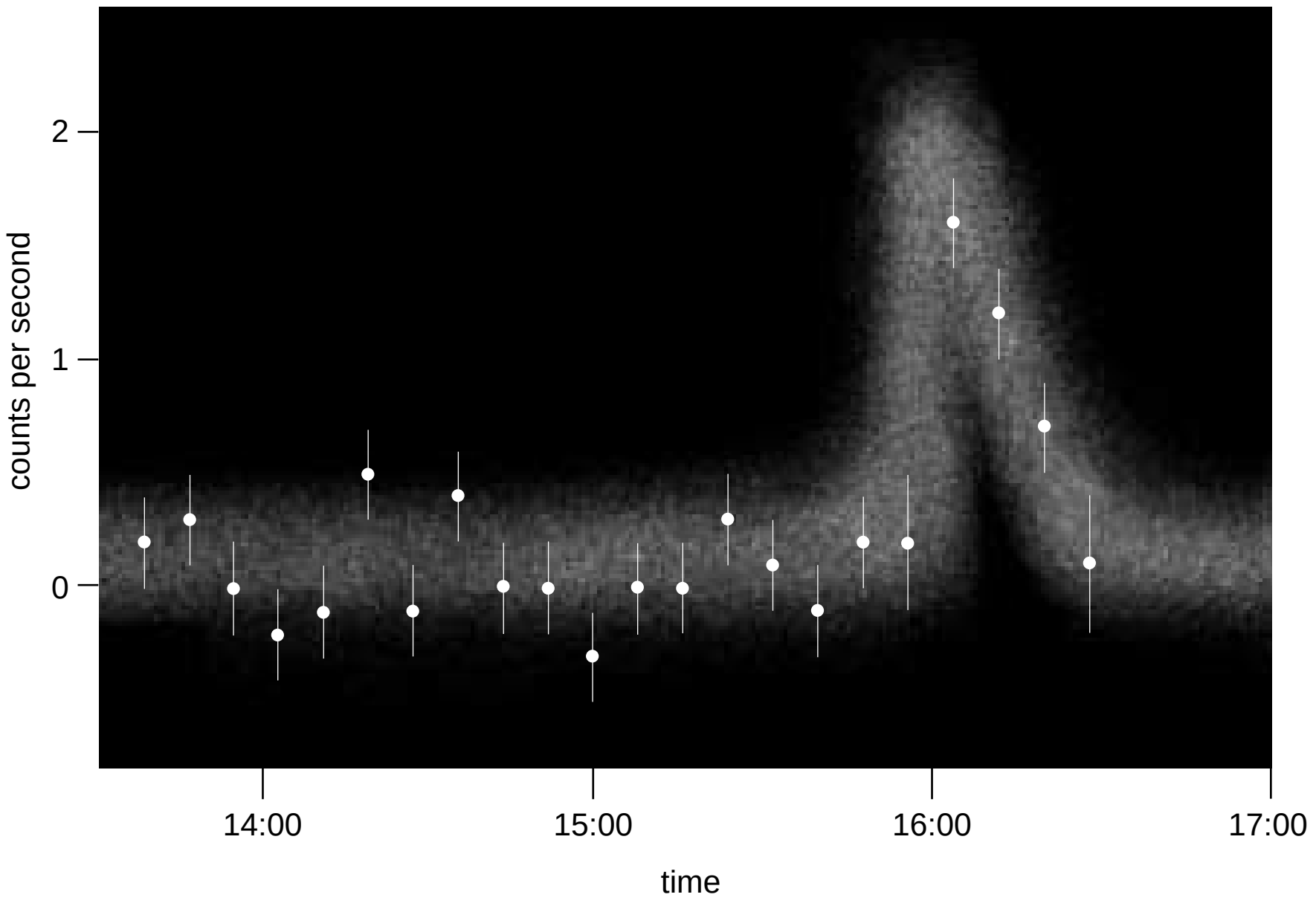

**Figure 5. Curious flaring of the star Pi[1] Ursae Majoris can be found in a series of x-ray measurements taken on January 31, 1984, using the orbiting observatory EXOSAT. Counts registered by this detector remain near zero for most of the record, then at 16:00 they rise abruptly and slowly fall to background levels. (After Landini *et al.*, 1986.)**

## The Superflare Riddle

Clearly these observations call for a new class of variables—sunlike stars that look very normal most of the time and then without warning give off an immense blast of energy. What could possibly trigger these sporadic superflares? This question drew me into a collaboration with Schaefer in 1998. At that time, Schaefer had just completed a series of careful spectroscopic measurements of the nine superflaring stars, but still he was baffled by what was causing the outbursts. So he decided to take advantage of a longstanding tradition: Every Thursday at noon the astronomy faculty and postdocs at Yale go off together and eat pizza.

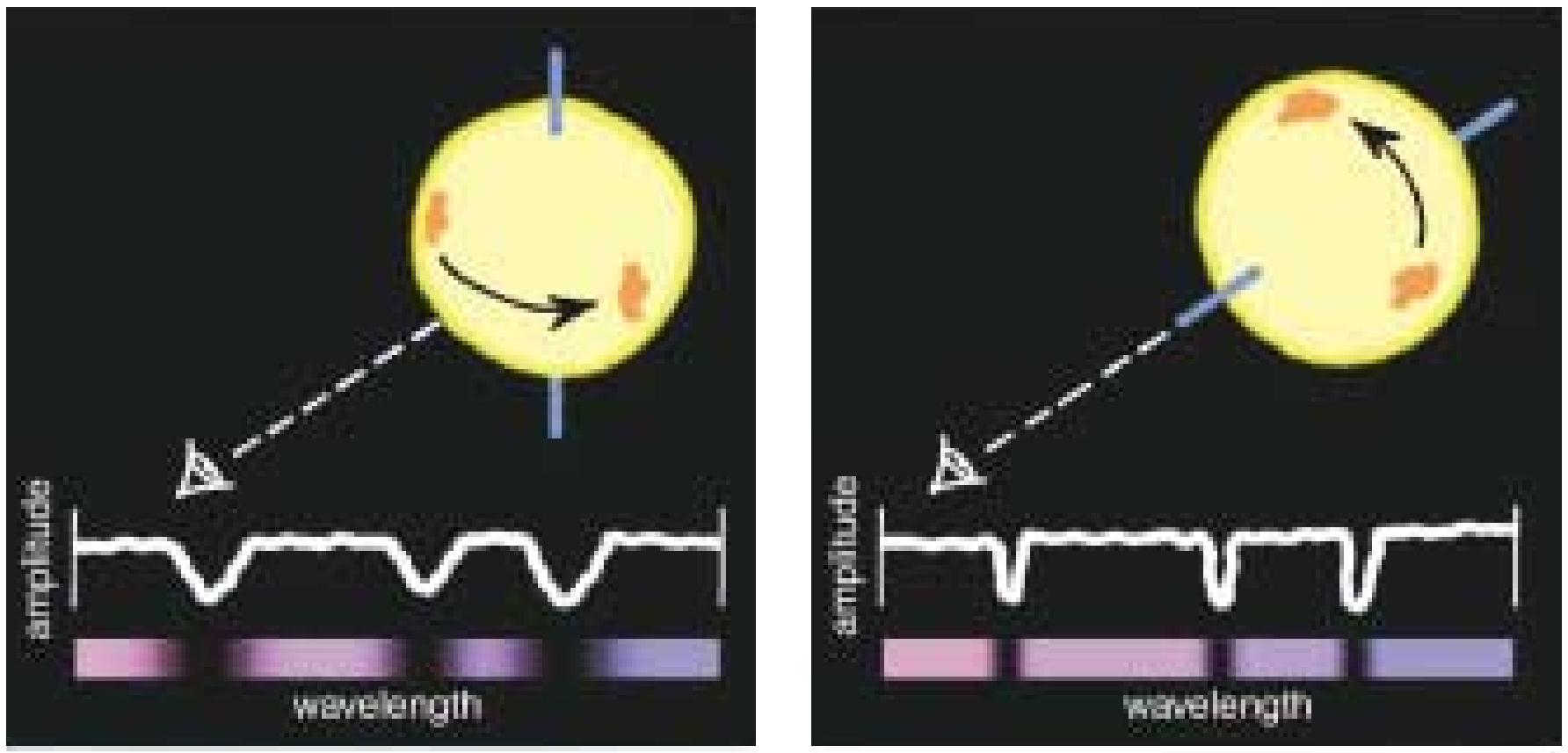

**Figure 6. Rotation of a distant star can be detected by changes in spectral lines. Light given off by a small patch on the surface of the spinning star will be shifted toward the blue end of the spectrum when the rotation moves that zone toward the observer. Similarly, light from that patch will be Doppler-shifted toward the red when rotation moves the source away. Because telescopic observations integrate light from all points on the face of the spinning star, astronomers detect only an overall broadening of spectral lines *(left)*. If, however, the rotation axis of the star is aligned closely with the line of sight, there is little motion toward or away from the observer and little widening in the spectral lines *(right)*. This method of measuring rotation thus gives only a minimum estimate of the rate.**

This outing is largely social, but science usually dominates the conversation. So one November afternoon, as we strolled slowly from our offices to the restaurant, Schaefer began systematically at the back of the line of astronomers and by ones and twos asked for ideas that might help solve the riddle of the superflares. Walking near the front, I heard him slowly overtaking my little group as he dismissed one proposal after another for various reasons. "No, they aren't supernovae, because the time scale and energy is all wrong," he might have said. "No, they aren't cataclysmic variables because they aren't close binaries," and so forth. Finally, by the time he was just a few steps behind me, I had heard the refrain enough times that I could repeat to myself the four most important constraints: The energy of a superflare is between $10^{33}$ and $10^{38}$ ergs; the duration varies from a fraction of an hour to days; the radiation comes out at wavelengths that span from at least the near infrared to x rays; and the Sun has not had any superflares recently, nor has a big one scorched the solar system within the last three billion years or so.

I thought to myself that these properties sounded quite a bit like some highly active binaries I used to study, which are named for the prime example: RS Canus Venaticorum, or RS CVn for short. These systems contain two stars that orbit in such close proximity that they have a period of between one and 15 days. The small separation means that the stars raise large tides on each other. This deformation causes the rotation of both stars to become synchronized to the orbital period of the pair. The resulting rapid spin and convection within the star generate powerful magnetic fields (a process called *dynamo action*), which in turn produce tremendous amounts of stellar activity in their chromospheres.

The RS CVn stars often have dark spots (similar to sunspots) covering a few tenths of their surfaces—a sign of strong magnetic fields—and they continually give off small-scale flares. In addition to this enhanced but essentially normal flaring, RS CVn stars also exhibit sporadic flares of immense proportions. Astrophysicists have proposed several explanations for these giant flares. What the different models have in common is that the energy of the flare is released suddenly by a phenomenon called *magnetic reconnection*.

Magnetic reconnection takes place after magnetic-field lines become tangled, pulled, stretched and twisted, and then suddenly reorganize into a simpler geometry, which stores far less energy in the magnetic field itself. This change is analogous to a bundle of elongated rubber bands suddenly snapping and releasing all of their pent-up energy at once. But magnetic-field lines, no matter how stretched, cannot break outright. The best they can do is to rearrange themselves pair-wise, splitting and simultaneously reconnecting so that they are no longer crossed. And whereas snapped rubber bands can fly off and hit things, reconnecting field lines do not separate from the object to which they are anchored. Rather, they pump energy into the ionized gas, the plasma, in which they reside.

The physics of magnetic reconnection is well studied but poorly understood. Investigators can observe it taking place in the laboratory, where it sometimes happens unexpectedly and releases enough energy to destroy expensive equipment. Magnetic reconnection is also clearly recognized to contribute to the production of flares and prominences on the surface of the Sun.

Astrophysicists believe that the industrial-strength flares on RS CVn binaries are most likely caused by the magnetic reconnection of field lines threading from one star to the other. In this situation, the entire volume enclosed by the great looping fields of both stars, a region perhaps 20 or more stellar radii across, is suddenly heated to millions of degrees when the lines reconnect. This event violently accelerates the ions and electrons in the stellar wind, producing broadband radiation and sending a burst of charged particles outward at nearly the speed of light.

With these thoughts displacing my craving for pizza, I turned around and mentioned to Schaefer that superflares seemed to have the same properties as the large events sometimes seen on RS CVn binaries. He immediately dismissed my idea—“These aren’t binaries.” But as I walked on, my mind kept churning away on the many similarities, and I became more and more convinced: If it looks like a duck, walks like a duck, sounds like a duck and blows up like a magnetic reconnection event, it probably *is* a reconnection event.

I was, however, troubled by one thing. With no stellar companion, the space in which a reconnection event

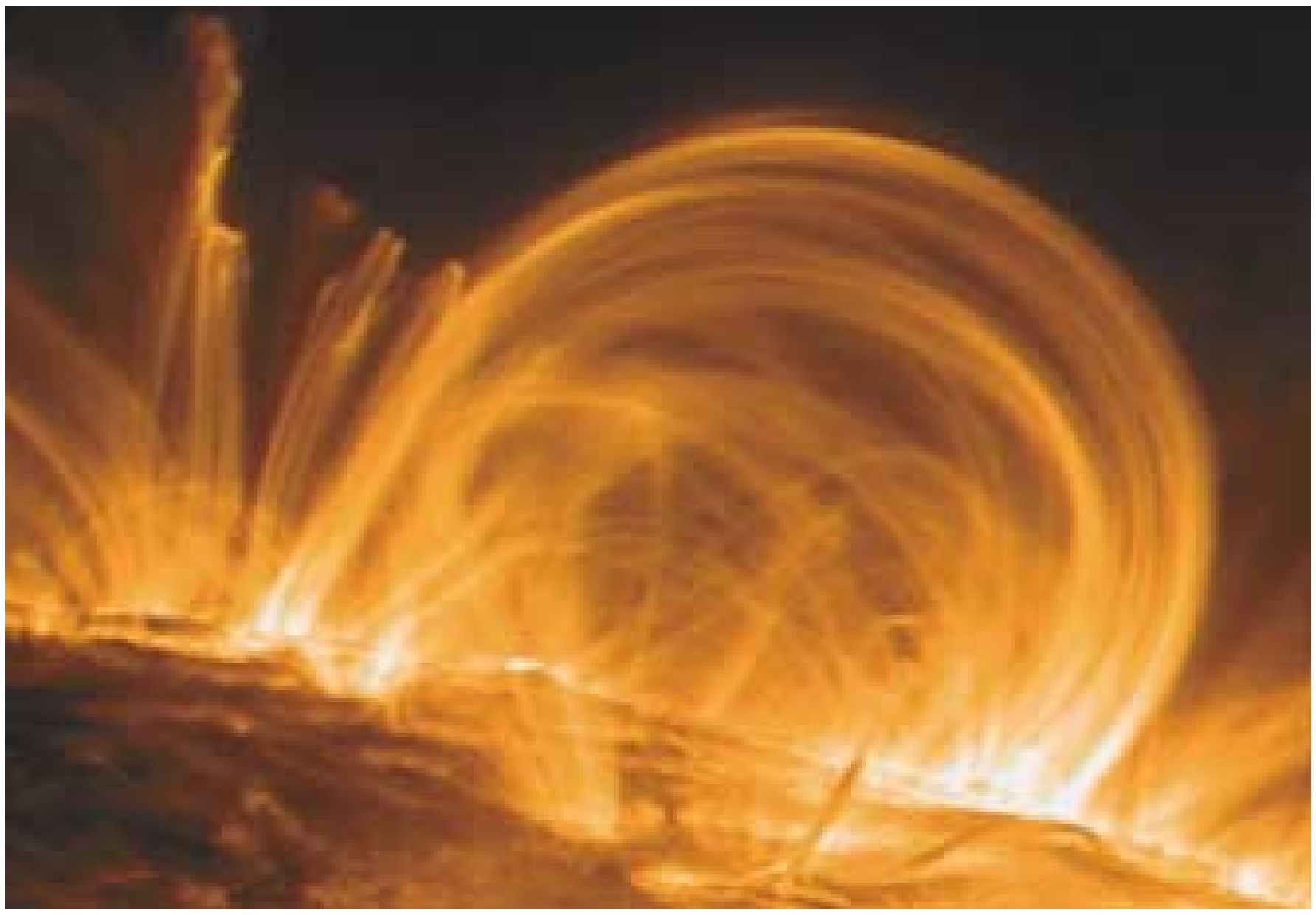

NASA

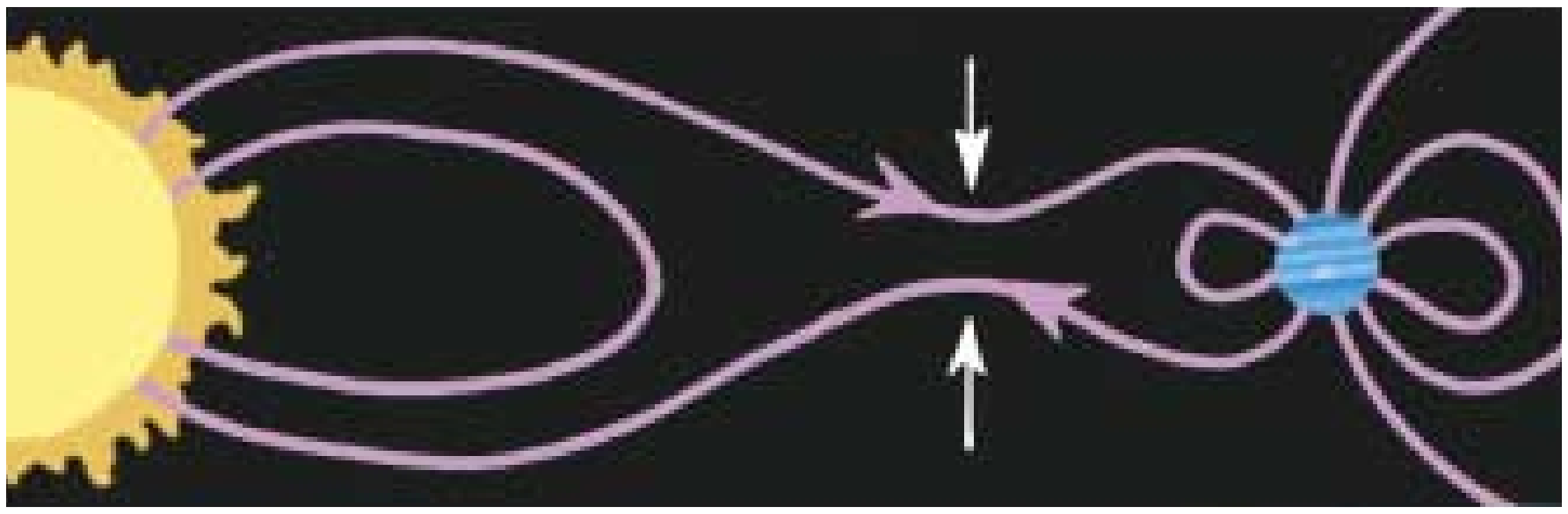

a

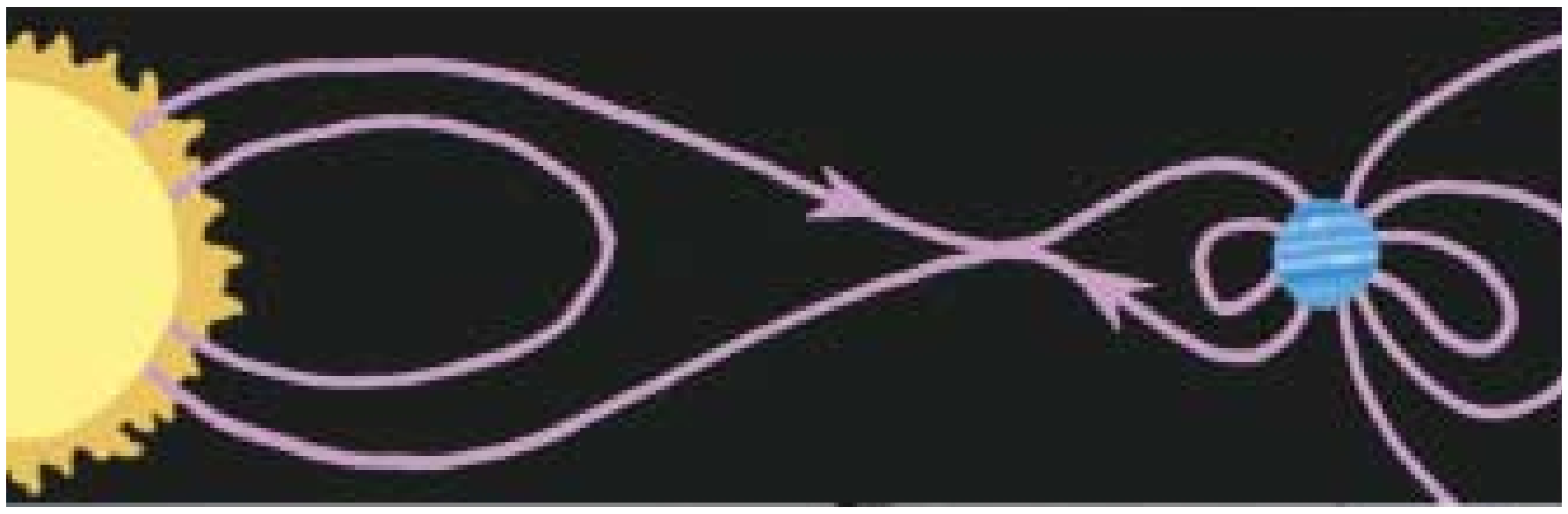

b

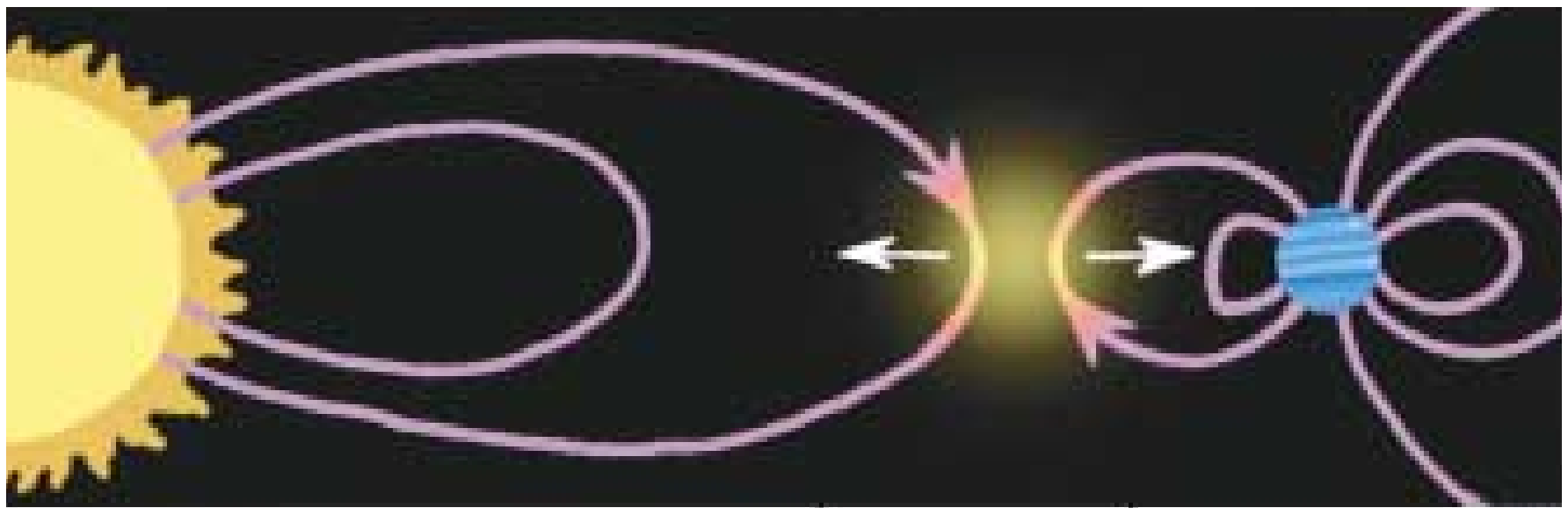

c

**Figure 7. Filaments of hot coronal gas reveal looping solar magnetic-field lines in an image obtained from the Transition Region and Coronal Explorer (TRACE) spacecraft *(top)*. The magnetic fields around stars with giant planets in close orbit are likely to be even more complicated, with some lines connecting the two bodies *(a)*. At times motion of the planet or of the ionized gases in space *(white arrows)* can force two fields lines together, allowing them to break from one configuration and reconnect in another *(b)*. This process injects energy into the surrounding plasma, accelerating charged particles and giving off a burst of high-frequency radiation *(c)*.**

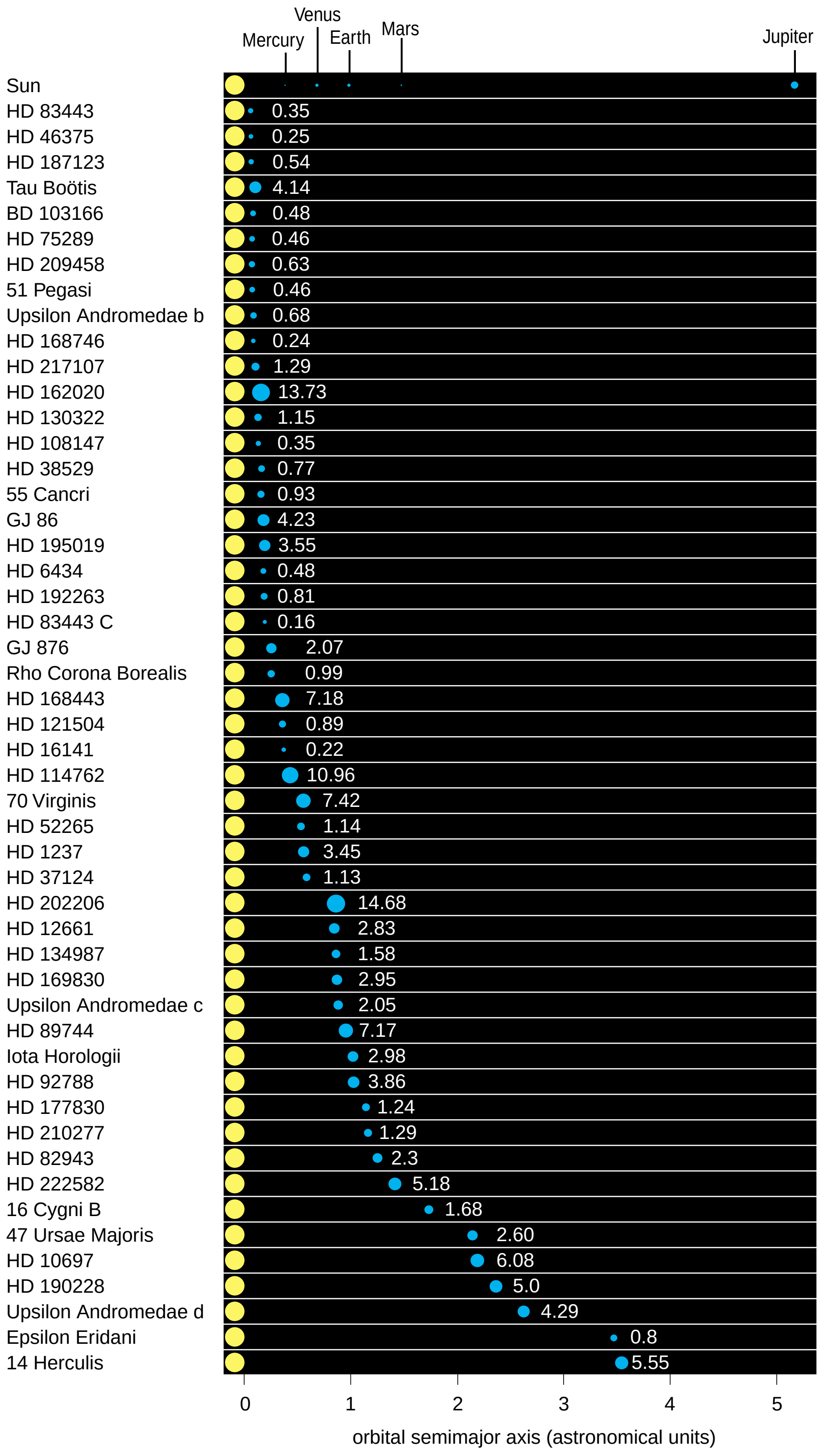

**Figure 8. Partial roster of suspected extrasolar planets demonstrates that massive, Jupiter-like bodies can circle in tight orbits. These discoveries have been made during the past five years from spectroscopic measurements that reveal wobbles in the parent star toward and away from Earth. Minimum estimates for the mass of each object are given here as the number of Jupiter masses. The diameter of the dot shows the relative size of these bodies, assuming that they have masses close to the values indicated and are of similar density to Jupiter. Such bodies probably support large magnetic fields, which may become entangled with the magnetic field of the parent star and spark a superflare.**

could happen would be little more than the size of the star, as is the case with solar flares. Because the energy released is proportional to the volume of the flaring region, a star-sized reconnection could never generate enough oomph to explain superflares. A second object with a magnetic field was clearly needed to make my idea work. Yet a second star in orbit, even one that was so small and dim that it could not be seen, would induce enough of a wobble in the bigger star to generate telltale Doppler shifts in its spectral lines. And despite having made careful spectroscopic observations of these stars, Schaefer had not found any such evidence for an unseen companion. Then in a heartbeat I thought of one such object that he would have missed: Jupiter.

Jupiter is only about a thousandth as massive as the Sun. Yet this gaseous planet has a strong magnetic field, which interacts with the solar wind. Indeed, astrophysicists believe that episodes of magnetic reconnection can explain some of the bursts of radio waves and higher-energy radiation that Jupiter occasionally emits.

So as we continued toward our destination, I imagined what would happen if Jupiter orbited the Sun closely enough that the magnetic fields of these two bodies could become fully entwined. A few steps later it struck me that Jupiter-like planets *have* been detected orbiting very close to their host stars. Astronomers now know of more than a few such planets, sometimes dubbed "hot Jupiters." Like the real Jupiter, they could well have large magnetic fields and thus could play the supporting role of unseen magnetic companion. The reconnection of field lines between such a planet and its star would then be expected to do just what it does on an RS CVn binary—generate immense flares now and again.

With his "These are not binaries" still lingering in the air, I strolled over to Schaefer and offered him my solution to the conundrum. We spent the rest of the walk (and more than a few slices of pizza) with him poking at the idea and trying, unsuccessfully, to find a problem with it. By the end of the meal, he agreed that the notion was quite compelling: It required no new physical process; it did not invoke any exotic type of astrophysical object; and it satisfied all of the observations. What is more, this explanation predicts that the stars involved should have rela-tively strong magnetic fields. Although estimates of stellar magnetic fields are difficult to make, two of the nine superflaring stars have been measured in this way, and both show high field strengths (in excess of a kilogauss) over substantial portions of their surfaces.

What is needed now is some direct evidence for giant planets in close orbit around these stars. Such determinations are, however, extremely challenging, which explains why only a few teams of astronomers have succeeded in discovering extrasolar planets. Geoff Marcy, who leads such an effort at the University of California, Berkeley and San Francisco State University, has surveyed one of the nine superflaring stars (Kappa Ceti) for planets and has found nothing. But he reports that the noise in these measurements is so large that he cannot rule out a planet the size of Saturn orbiting within 0.1 astronomical unit (or A.U., about 150 million kilometers) of the star—or even a Jupiter-like planet circling somewhat farther out. So Schaefer and I will have to wait for more definite word from Marcy and his fellow planet-hunters before we know for sure whether we are right.

**Spectacular Fireworks**

Even without conclusive proof of the idea, it is tempting to think that a mechanism for generating superflares is now known, at least in outline. It is also tempting to try to envision what one of these strange planetary systems looks like up close. There must be at least one giant planet orbiting searingly near to the central yellow star—perhaps more closely than Mercury orbits the Sun. The planet rotates rapidly, a prerequisite for generating a large magnetic field, and this fast spin sets up a conspicuous pattern of banding in its atmosphere. The large striped planet, and perhaps its several moons, are probably bathed in x rays and ultraviolet light produced by the stellar corona and by the interplay of strong magnetic fields. Some of the field lines of this hot Jupiter connect fully with those emerging from the star, permitting charged particles from the stellar corona to rain down on the planet and its companions, lighting up the nighttime sky with bright aurora.

Over the years, the orbit of this planet has slowly tangled its magnetic field with that of the star. Like the spiral spring of a clock, the linked field lines are wound around and around. Now is the moment they break loose. In just minutes, all this potential energy courses into the surrounding stellar plasma, rapidly accelerating many of these charged particles. The affected electrons and ions immediately give off a blast of x rays and ultraviolet radiation. Some fly outward at nearly the speed of light.

Is this the day that some extraterrestrial civilization meets its end? Probably not. But in time someone on Earth may look up at the night sky and notice a seemingly normal, sunlike star that for a few minutes appears substantially brighter than it ever has before. To my mind, this is a good day.

**Bibliography**

Ashbrook, J. 1959. The S Fornacis puzzle. *Sky and Telescope* 18:427.

Beardsley, W. R., G. Gatewood and K. W. Kamper. 1974. A study of an early flare, radial velocities, and parallax residuals for possible orbital motion of HD 103095 (Groombridge 1830). *The Astrophysical Journal* 194:637–643.

Klimchuk, J. A. 1997. Magnetic energy release on the Sun. *Nature* 386:760–761.

Landini, M., C. Monsignori Fossi, R. Pallavicini and L. Piro, 1986. EXOSAT detection of an X-ray flare from the solar type star $\pi^1$ UMa. *Astronomy and Astrophysics* 157:217–222.

Rubenstein, E. P., and B. E. Schaefer. 2000. Are superflares on solar analogues caused by extrasolar planets? *The Astrophysical Journal* 529:1,031–1,033.

Schaefer, B. E. 1989. Flashes from normal stars. *The Astrophysical Journal* 337:927–933.

Schaefer, B. E., J. R. King and C. P. Deliyannis. 2000. Superflares on ordinary solar-type stars. *The Astrophysical Journal* 529:1,026–1,030.

Links to Internet resources for further exploration of "Superflares and Giant Planets" are available on the *American Scientist* Web site:

http://www.americanscientist.org/articles/01articles/rubenstein.html

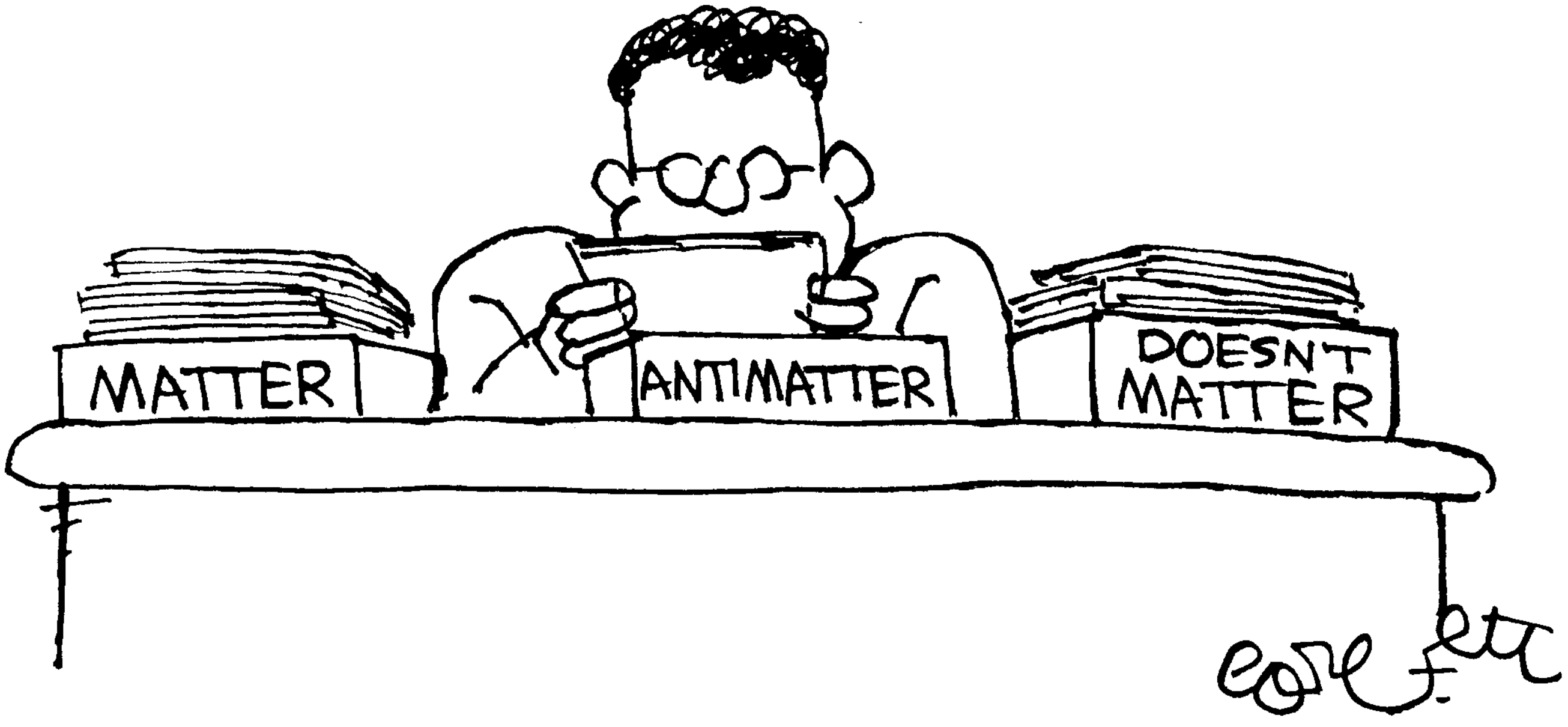